\documentclass[useAMS,usenatbib]{mn2e}
\usepackage{epsfig}
\usepackage{color}
\usepackage{subfigure}
\usepackage{natbib}
\usepackage{aas_macros}

\title[Flexion and cluster mass]{Improving lensing cluster mass estimate with flexion}

\author[V.F. Cardone et al.]{V.F. Cardone$^{1}$\thanks{Corresponding author\,: {\tt winnyenodrac@gmail.com}}, M. Vicinanza$^{1,2,3}$, X. Er$^{1}$, R. Maoli$^{3}$, R. Scaramella$^{1}$ \\
$^1$I.N.A.F.\,-\,Osservatorio Astronomico di Roma, via Frascati 33, 00040 - Monte Porzio Catone (Roma), Italy \\
$^2$Dipartimento di Fisica, Universit\`{a} di Roma "Tor Vergata", via della Ricerca Scientifica 1, 00133 - Roma, Italy \\
$^3$Dipartimento di Fisica, Universit\`{a} di Roma "La Sapienza", Piazzale Aldo Moro, 00185 - Roma, Italy \\}

\date{Accepted xxx, Received yyy, in original form zzz}

\begin{document}

\maketitle

\begin{abstract}

Gravitational lensing has long been considered as a valuable tool to determine the total mass of galaxy clusters. The shear profile as inferred from the statistics of ellipticity of background galaxies allows to probe the cluster intermediate and outer regions thus determining the virial mass estimate. However, the mass sheet degeneracy and the need for a large number of background galaxies motivate the search for alternative tracers which can break the degeneracy among model parameters and hence improve the accuracy of the mass estimate. Lensing flexion, i.e. the third derivative of the lensing potential, has been suggested as a good answer to the above quest since it probes the details of the mass profile. We investigate here whether this is indeed the case considering jointly using weak lensing, magnification and flexion. We use a Fisher matrix analysis to forecast the relative improvement in the mass accuracy for different assumptions on the shear and flexion signal\,-\,to\,-\,noise (S/N) ratio also varying the cluster mass, redshift, and ellipticity. It turns out that the error on the cluster mass may be reduced up to a factor $\sim 2$ for reasonable values of the flexion S/N ratio. As a general result, we get that the improvement in mass accuracy is larger for more flattened haloes, but it extracting general trends is a difficult because of the many parameters at play. We nevertheless find that flexion is as efficient as magnification to increase the accuracy in both mass and concentration determination.

\end{abstract}


\section{Introduction}

In a hierarchical bottom\,-\,up scenario, structures formation proceeds by merging of low mass objects into high mass systems. Being the most massive gravitationally bound structures in the universe, galaxy clusters emerge as the final point of this process thus making them ideal probes of the structure formation history. Both their density profile and mass function are intimately related to the underlying background cosmological model and the growth of structures thus offering the intriguing possibility to contrain both the cinematics and dynamics of the universe. In particular, they trace the exponential cutoff of the halo mass function thus being particularly sensitive to the details of the theory of gravity. Weighting galaxy clusters has therefore become a rewarding yet daunting business leading astronomers to look for different methods to achieve this difficult goal.

Weak gravitational lensing soon emerged as one of the most valuable candidate to end this quest \citep{BS2001,Ref2003,Sch2006,Mun2008}. Massive clusters bend the optical path of light rays causing a distortion and a magnification of the image of a background source so that a supposedly circular object looks like an elliptical one. Needless to say, sources are not circular so that the lensing effect can only be measured statistically thus asking for a large source number density. As difficult as this task is, it is nevertheless possible to use this effect to trace the shear profile of the clusters as first shown by \cite{Ty90} measuring the statistical apparent alignment of galaxies due to the gravitational lensing. Since their initial measurements, different methods were proposed to convert these data into a cluster mass profile and hence a determination of its total mass (see, e.g., \citealt{KS93,SS2001}). Although mass measurement through weak lensing are now routinely performed, the situation is far from being problems free. Indeed, shear profile only traces the cluster mass distribution in the intermediate to outer regions thus being unable to constrain the concentration parameter $c_{vir} = R_{vir}/R_s$ with $R_{vir}$  $(R_s)$ the virial (scale) radius of the mass profile. The mass\,-\,concentration degeneracy then transfers part of the uncertainty on $c_{vir}$ on the mass $M_{vir}$ thus weakening the constraints on this latter quantity. Moreover, it is not always possible to get a sufficiently high source number density thus preventing from narrow down the range for the mass. As a possible way out of both these problems, one has to add a second tracer to shear profile with strong lensing as the most common candidate. Indeed, lensing in the strong regime traces the inner regions thus pushing down the statistical error on $c_{vir}$ and hence on the mass. Unfortunately, strong lensing features are difficult to detect and their modeling is strongly sensitive to the presence of substructures whose amount and distribution are far from being well known.

An alternative probe is represented by gravitational lensing flexion defined as the third derivative of the lensing potential \cite{GN02,GB05,Betal06}. Spin\,-\,1 and spin\,-\,3 flexion terms are responsible for the mapping of intrinsically round source into off centred and arc\,-\,like images thus adding further features to the lensing phenomena. Flexion can be measured from higher order shape moments \citep{Ok2007} or from image model fitting \citep{Ref12003,RB2003,Cain2011} thus allowing to probe small scale features of the mass profile as first demonstrated by \cite{Leo07} for Abell\,1689. As a bonus point, it is also expected that third order galaxy moments are much smaller than second order ones. As a consequence, the dispersion of intrinsic flexion should be much smaller than the intrinsic ellipticity one thus lowering the shape noise which severely downgrade the potential of the shear method. 

Following preliminary analysis, \cite{Er12} have investigated how adding flexion to shear improves the cluter mass determination for two different clusters using mock data. However, the difficulty in simulating data prevent them to explore both the mass\,-\,redshift parameter space and the assumptions on the shear and flexion S/N. We therefore extend their analysis here by reverting to a Fisher matrix analysis. This makes it possible to investigate how the constraints on the mass changes as function of both the mass itself and the cluster redshift. Moreover, we also investigate the dependence on the axial ratio. Since measurin flexion is a daunting task yet to be satisfactorily achieved, we investigate how the results change as a function of the shear and flexion S/N in order to examine which strategy is more rewarding from the point of view of getting a significant boost in mass accuracy.

The plan of the paper is as follows. Basic lensing quantities and the relevant formulae for the particular cluster model adopted are summarized in Sect.\,2, while Sec.\,3 presents the mass determination methodology combining shear, magnification and flexion. Details on the computation of the Fisher matrix are given in Sect.\,4, while Sect.\,5 presents the main results of the paper. Conclusions are then summarized in Sect.\,6, while in the Appendix we qualitatively address the impact of substructures.

\section{Lensing basics}

The image deformation due to the lensing effect of a cluster along the line of sight to a galaxy source depends on the lens surface density profile $\Sigma({\bf x})$ under the usual thin lens approximation. In particular, the isotropic focusing of light rays is determined by the convergence\footnote{Hereafter, we will assume spherical symmetry so that all the relevant quantities will only depend on the magnitude $\theta$ of the vector position on the plane.} $\kappa(\theta) = \Sigma(\theta)/\Sigma_{crit}(z_l, z_s)$ where the critical surface density reads

\begin{equation}
\Sigma_{crit}(z_l, z_s) = \frac{c^2 \beta^{-1}(z_l, z_s)}{4 \pi G D_l}
\label{eq: sigmacrit}
\end{equation}
with $(z_l, z_s)$ the lens and source redshifts, $\beta = D_{ls}/D_s$ and $(D_l, D_s, D_{ls})$ the angular diameter distances from observer to lens, observer to source, lens to source, respectively.

While the convergence $\kappa$ causes isotropic distortions, the shear $\gamma(\theta)$ induces a quadrupole deformation which can be statistically observed from the ellipticity of a large number of background sources. The tangential shear at an angular distance $\theta$ is given by \citep{K95}

\begin{equation}
\gamma_{+}(\theta) = \kappa(< \theta) - \kappa(\theta)
\label{eq: gammaplus}
\end{equation}
while the $45^o$ rotated component is identically null. In Eq.(\ref{eq: gammaplus}), $\kappa(\theta)$ is the azimuthally averaged convergence at the radius $\theta$ and $\kappa(<\theta)$ is its average value within $\theta$. What can be inferred from observations is, actually, not the shear itself, but the reduced shear defined as

\begin{equation}
g(\theta) = \frac{\gamma(\theta)}{1 - \kappa(\theta)} \ ,
\label{eq: redshear}
\end{equation}
which reduces to $\gamma$ in the weak lensing limit $\kappa, \gamma << 1$. In cluster lensing, this regime is approximately verified all over the radial profile but in the very inner regions. Although the data typically probe well outside the centre, we will nevertheless take care of the difference between $g$ and $\gamma$.

The reduced shear is invariant under a linear transformation replacing $\kappa(\theta)$ with $\lambda \kappa(\theta) + (1 - \lambda)$ and $\gamma(\theta)$ with $\lambda \gamma(\theta)$, $\lambda$ being an arbitrary constant \citep{SS95}. In order to break this degeneracy (typically referred to as the mass sheet degeneracy), one can rely on the magnification, given by

\begin{equation}
\mu(\theta) = \frac{1}{[1 - \kappa(\theta)]^2 - \gamma^2(\theta)} \ ,
\label{eq: defmag}
\end{equation}
which transforms as $\lambda^2 \mu(\theta)$. The magnification alters the number counts of background galaxies causing a depletion or an increase with respect to the unlensed case depending on the effective slope of the source number density with luminosity \citep{B95}. This offers a different tool to constrain the lens properties as we will see later.

Both the convergence and the shear can be expressed in terms of the second derivative of the lensing potential. Going to the next order, one gets the flexion \citep{GN02,GB05,Betal06}. There are actually two different kind of flexion, namely the spin\,-\,1 first flexion

\begin{equation}
{\cal{F}}(\theta) = \nabla \kappa(\theta) \ ,
\label{eq: firstflex}
\end{equation}
and the spin\,-\,3 second flexion

\begin{equation}
{\cal{G}}(\theta) = \nabla \gamma(\theta) \ ,
\label{eq: secondflex}
\end{equation}
where $\nabla = \partial_1 + {\rm i} \partial_2$ is the complex gradient. These two indepedent fields specify the {\it arciness} of the distorted image causing a centre\,-\,shift and  an arc\,-\,like distortion, respectively, possibly leading to a banana\,-\,like shape in extreme conditions. What makes them interesting tools is their being a probe of the local gradient of the convergence and shear fields thus making it possible to better constrain the lens projected density profile.

\subsection{NFW halo}

The key ingredient for the computation of the convergence, shear and flexion is the lens density profile. We are here interested in cluster mass determination so that we model the lens a single dark matter halo hence neglecting both substructures and the central brightest cluster galaxy (BCG). Note that, since we will use data probing the outer regions, the BCG gives indeed a fully negligible contribution. We then describe the cluster density profile\footnote{It is worth noting that we are here neglecting the contribution of substructures so that we can use a smooth profile to compute the lensing quantities. In Appendix A we put down a simple yet realistic model to evaluate the systematics introduced by this approximation. As can be naively expected, neglecting substructures has no impact on the shear profile, but can be significant for flexion since this is more sensible to local perturbations. However, the systematics errors induced by neglecting substructure is everywhere smaller than the statistical one so that we are confident that a smooth profile is enough for our aims.} using the popular NFW model \citep{NFW96,NFW97}

\begin{equation}
\rho(r) = \rho_s (r/r_s)^{-1} (1 + r/r_s)^{-2}
\label{eq: rhonfw}
\end{equation}
with $(\rho_s, r_s)$ scaling quantities which are typically replaced by two alternative parameters, namely the virial mass $M_{vir}$, defined as the mass within the radius enclosing a mean density equals to $\Delta_{vir} \rho_{crit}(z)$, and the concentration $c_{vir} = R_{vir}/r_s$, with $R_{vir}$ the virial radius. We follow \cite{BN98} to set the virial overdensity $\Delta_{vir}(z)$ and take as fiducial cosmology a flat $\Lambda$CDM model with $(\Omega_M, h) = (0.306, 0.678)$ in agreement with recent Planck results \citep{PlanckXIII}.

The lensing properties of the NFW model may be analytically expressed if the halo is spherically symmetric \citep{Bart96,WB00,Betal06}. However, this is not the case in realistic applications so that we will allow for a possible flattening quantified by the axial ratio $q$ and replacing $r$ with $r/(1 - \varepsilon)$ with $\varepsilon = (1 - q)/(1 + q)$ in Eq.(\ref{eq: rhonfw}). The computation of the lensing properties now demands for numerical integrations, but it still manageable. Adopting the complex notation, it is \citep{K01,HB09} 

\begin{equation}
\kappa = \frac{1}{2} (\psi_{11} + \psi_{22}) \ ,
\label{eq: ellconv}
\end{equation}

\begin{equation}
\gamma = \frac{1}{2} (\psi_{11} - \psi_{22}) + {\rm i} \psi_{12} \ , 
\label{eq: ellgamma}
\end{equation}

\begin{equation}
{\cal{F}} = \frac{1}{2} [ \psi_{111} + \psi_{122} + {\rm i} (\psi_{112} + \psi{222}) ] \ ,
\label{eq: elleffe}
\end{equation}

\begin{equation}
{\cal{G}} = \frac{1}{2} [ \psi_{111} - 3 \psi_{122} + {\rm i} (\psi_{112} - \psi_{222}) ] \ ,
\label{eq: ellgi}
\end{equation}
where $\psi$ is the lensing potential and labels denote partial differentiation. Derivatives in the image plane $(x, y)$ are given by

\begin{displaymath}
\psi_{11} = 2 q x^2 K_0 + q J_0  \ ,
\end{displaymath}

\begin{displaymath}
\psi_{22} = 2 q y^2 K_0 + q J_1 \ ,
\end{displaymath}

\begin{displaymath}
\psi_{12} = 2 q x y K_1 \ ,
\end{displaymath}

\begin{displaymath}
\psi_{111} = 6 q x K_0 + 4 q x^3 L_0 \ ,
\end{displaymath}

\begin{displaymath}
\psi_{222} = 6 q y K_2 + 4 q y^3 L_3 \ ,
\end{displaymath}

\begin{displaymath}
\psi_{112} = 2 q y K_1 + 4 q x^2 y L_1 \ ,
\end{displaymath}

\begin{displaymath}
\psi_{122} = 2 q x K_1 + 4 q y^2 x L_2 \ ,
\end{displaymath}
having defined

\begin{displaymath}
J_n(x, y) = \int_{0}^{1}{\frac{\kappa(\xi^2) du}{[1 - (1 - q^2) u]^{n+1/2}}} \ ,
\end{displaymath}

\begin{displaymath}
K_n(x, y) = \int_{0}^{1}{\frac{u \kappa^{\prime}(\xi^2) du}{[1 - (1 - q^2) u]^{n+1/2}}} \ ,
\end{displaymath}

\begin{displaymath}
L_n(x, y) = \int_{0}^{1}{\frac{u^2 \kappa^{\prime \prime}(\xi^2) du}{[1 - (1 - q^2) u]^{n+1/2}}} \ ,
\end{displaymath}
where the prime denotes differentiation with respect to the auxiliary variable

\begin{displaymath}
\xi^2(u) = u \left [  x^2 + \frac{y^2}{1 - (1 - q^2) u} \right ] \ .
\end{displaymath}
Inserting the NFW convergence profile into the above formulae and making some boring but easy algebra, one finally gets the following short hand results for the the magnitude of the quantities of interest

\begin{equation}
\kappa(\xi, \eta) = (q \kappa_s/2) \tilde{\kappa}(\xi, \eta, c_{vir}, q) \ ,
\label{eq: kappanfwell}
\end{equation}

\begin{equation}
\gamma(\xi, \eta) = (q \kappa_s/2) \tilde{\gamma}(\xi, \eta, c_{vir}, q) \ ,
\label{eq: gammanfwell}
\end{equation}

\begin{equation}
{\cal{F}}(\xi, \eta) = (q \kappa_s/2 r_s) \tilde{{\cal{F}}}(\xi, \eta, c_{vir}, q) \ ,
\label{eq: effenfwell}
\end{equation}

\begin{equation}
{\cal{G}}(\xi, \eta) = (q \kappa_s/2 r_s) \tilde{{\cal{G}}}(\xi, \eta, c_{vir}, q) \ ,
\label{eq: ginfwell}
\end{equation}
with $(\xi, \eta) = (x, y)/\theta_s$ dimensionless coordinates in the image plane, $\theta_s = r_s/D_{l}$, and $\kappa_s$ a scaling value given by

\begin{equation}
\kappa_s = \frac{M_{vir}/4 \pi R_{vir}^2}{\Sigma_{crit}(z_l, z_s)}
\frac{c_{vir}^2}{\ln{(1 + c_{vir})} - c_{vir}/(1 + c_{vir})} \ .
\label{eq: kappasdef}
\end{equation}
The tilted functions are a combination of the numerical integrals entering the derivatives of the lensing potential that we do not report here for sake of brevity. We, however, stress that they depend on the concentration $c_{vir}$ and the axial ratio $q$, but not on the virial mass $M_{vir}$ which only enters through $\kappa_s$ and $\kappa_s/r_s$.

A final remark is in order. As can be read from Eqs.(\ref{eq: kappanfwell})\,-\,(\ref{eq: ginfwell}), the lensing quantities depend on both coordinates in the image plane or, put in other words, on both the polar coordinates $(R, \varphi)$. On the contrary, the observed counterparts are not given as a function of $(R, \varphi)$, but only as measured at a given distance $R$ from the cluster centre. We therefore compare observations with the corresponding angle averaged theoretical quantity defined as 

\begin{equation}
{\cal{O}}(R, q) = \frac{1}{2 \pi} \int_{0}^{2 \pi}{{\cal{O}}(R, \varphi, q) d\varphi}
\label{eq: angav}
\end{equation}
with $(\xi, \eta) = (R/r_s) (\cos{\varphi}, \sin{\varphi} )$ and ${\cal{O}} = (\kappa, \gamma, {\cal{F}}, {\cal{G}})$. 

\section{Cluster mass from lensing}

The reduced shear can be measured from the ellipticity of the background galaxies and can be used to constrain the cluster mass profile and hence determine its mass. Actually, one should also take into account the distribution in redshift of the sources which makes the convergence dependent on the background galaxies. To this end, we compare the estimated reduced shear $g_{obs}(\theta)$ with the theoretical value after correcting it for the nonlinearities induced by averaging over the source distribution. Following \cite{SS97,U13}, we approximate this latter as

\begin{equation}
g_{th}(\theta) = \frac{\langle W_ g \rangle \gamma_{\infty}(\theta)}{1 - f_W \langle W_g \rangle \kappa_{\infty}(\theta)}
\label{eq: gth}
\end{equation}
with $f_W = \langle W_g^2 \rangle/\langle W_g \rangle^2 \simeq 1$, $W_g = \beta(z_l, z_s)/\beta(z_l, z_s \rightarrow \infty)$, and we define

\begin{equation}
\langle W_g \rangle = \left [ \int_{0}^{\infty}{\beta(z) n_g(z) dz} \right ] \left [ \int_{0}^{\infty}{n_g(z) dz} \right ]^{-1} \ ,
\label{eq: wgdef}
\end{equation}
$n_g(z)$ being the source redshift distribution. Finally, in Eq.(\ref{eq: gth}), $\kappa_{\infty}$ and $\gamma_{\infty}$ are the convergence and the shear for a source at infinity, i.e. they are obtained by the above formulae setting $z_s \rightarrow \infty$ in Eq.(\ref{eq: kappasdef}).

We can now introduce the likelihood function for the shear data as (see, e.g., \citealt{Clash14})

\begin{equation}
\ln{{\cal{L}}_g({\bf p})} = -\frac{1}{2} \sum_{i = 1}^{{\cal{N}}_{wl}}{\left [ \frac{g_{obs}(\theta_i) - g_{th}(\theta_i, {\bf p})}{\sigma_g(\theta_i)} \right ]^2}
\label{eq: wllike}
\end{equation}
where $\sigma_g(\theta_i)$ is the error on the reduced shear measured from galaxies in a circular bin centred on $\theta_i$ and the sum is over the ${\cal{N}}_{wl}$ bins. The likelihood is a function of the halo model parameters $(\log{M_{vir}}, c_{vir}, q)$ and the nuisance parameter $f_W$.

As already hinted at above, the reduced shear is affected by the mass\,-\,sheet degeneracy which can be broken relying on magnification $\mu(\theta)$. As for the shear, it is not $\mu(\theta)$ which is determined from the data, but rather its impact on the source number density. It is indeed possible to show that the number of galaxies with redshift $z$ and magnitude less than a given threshold $m$ is given by \citep{B95}

\begin{equation}
N_{\mu}(\theta, z, < m) = \bar{N}_{\mu}(z, < m) [\mu(\theta)]^{2.5 s - 1}
\label{eq: nmuzed}
\end{equation}
where $\bar{N}_{\mu}$ is the unlensed mean source counts and $s = d\log{\bar{N}}_{\mu}(z, < m)/dm$ is the logarithmic slope of the unlensed source counts. In practical observations, the nonvanishing and unresolved small scale angular correlation can increase the variance of counts in cells thus washing out the lensing signal \citep{vW00}. This local clustering noise can be overcome by performing an azimuthal average around the cluster so that the observed quantity is now $n_{\mu}(\theta) = dN_{\mu}(\theta, z, < m)/d\Omega$ \citep{Uetal11,U13}. We can therefore define a likelihood function as

\begin{equation}
\ln{{\cal{L}}_{\mu}({\bf p})} = -\frac{1}{2} \sum_{i = 1}^{{\cal{N}}_{\mu}}{\left [
\frac{n_{\mu}^{obs}(\theta_i) - n_{\mu}^{th}(\theta_i, {\bf p})}{\sigma_{\mu}(\theta_i)} \right ]^2}
\label{eq: likemag}
\end{equation}
where the sum is over the ${\cal{N}}_{\mu}$ bins and, to take into account the source redshift distribution, the theoretical number density is computed as

\begin{equation}
n_{\mu}^{th}(\theta, {\bf p}) = \bar{n}_{\mu} \langle \mu^{-1}(\theta) \rangle^{1 - 2.5 s_{eff}}
\label{eq: nmuth}
\end{equation}
with $\bar{n}_{\mu}$ the unlensed mean surface density, $s_{eff}$ the effective logarithmic slope and

\begin{equation}
\langle \mu^{-1}(\theta) \rangle = [1 - \langle W_g \rangle \kappa_{\infty}(\theta) ]^2 -
\langle W_g \rangle^2 \gamma_{\infty}^2(\theta) \ 
\label{eq: muth}
\end{equation}
with $(\kappa_{\infty}, \gamma_{\infty})$ estimated taking care of the cluster ellipticity. The likelihood function defined above\footnote{Note that we have here assumed that the redshift distribution of the source galaxies used for the measurement of $n_{\mu}(\theta)$ is the same as the one of the galaxies used in the shear determination. Should this not be the case, Eq.(\ref{eq: muth}) still holds provided one replace $\langle W_g \rangle$ with $\langle W_{\mu} \rangle$ being this latter given by Eq.(\ref{eq: wgdef}) with $n_g(z)$ set to the actual redshift distribution. We will neglect this difference here since we are not interested in fitting real data, but only to explore combined constraints from shear and magnification.} is a function of the halo model parameters $(\log{M_{vir}}, c_{vir}, q)$ and the nuisance parameters $(\bar{n}_{\mu}, s_{eff})$. Although these latter quantities may be guessed from fields away from the cluster of interest, we will nevertheless include them in the list of unknown quantities and marginalize over them in the following analysis.

While the use of reduced shear and magnification to constrain the cluser mass profiles is a nowadays well established tool, our aim here is to add flexion in order to investigate whether and under which conditions this can help to reduce significantly the errors on the mass determination. To this end, we introduce two further likelihood functions defined as

\begin{equation}
\ln{{\cal{L}}_{{\cal{F}}}({\bf p})} = -\frac{1}{2} \sum_{i = 1}^{{\cal{N}}_F}{ \left [
\frac{{\cal{F}}_{obs}(\theta_i) - {\cal{F}}_{th}(\theta_i, {\bf p})}{\sigma_{{\cal{F}}}(\theta_i)} \right ]^2} \ ,
\label{eq: likefirstflex}
\end{equation}

\begin{equation}
\ln{{\cal{L}}_{{\cal{G}}}({\bf p})} = -\frac{1}{2} \sum_{i = 1}^{{\cal{N}}_G}{ \left [
\frac{{\cal{G}}_{obs}(\theta_i) - {\cal{G}}_{th}(\theta_i, {\bf p})}{\sigma_{{\cal{G}}}(\theta_i)} \right ]^2} \ ,
\label{eq: likefsecondflex}
\end{equation}
with obvious meaning of the symbols. As for the reduced shear, the theoretical quantities are defined after averaging over the source redshift distribution thus leading to ${\cal{F}}_{th}(\theta) = \langle W_g \rangle {\cal{F}}_{\infty}(\theta)$ with ${\cal{F}}_{\infty}$ the flexion for a source at infinity. A similar formula holds for the second flexion ${\cal{G}}(\theta)$.

It is worth noting that, as for the shear, the flexion is actually not directly observable. Indeed, what is measured from the data is \citep{SE08}

\begin{equation}
{\cal{F}}_{meas}(\theta) = \frac{{\cal{F}}(\theta) + g(\theta) {\cal{F}}^{\star}(\theta)}{1 - \kappa(\theta)}
\label{eq: firstflexobs}
\end{equation}
where the flexion is here considered as a complex number and ${\cal{F}}^{\star}(\theta)$ is the complex conjugate. ${\cal{F}}_{meas}(\theta)$ reduces to ${\cal{F}}(\theta)$ in the weak lensing limit $g(\theta), \kappa(\theta) << 1$. Although over the radial range probed by the data the difference is far smaller than the typical uncertainty, we nevertheless take care of it in the following when comparing the theoretical flexions with the measured ones.

The full likelihood can now be naively written as

\begin{eqnarray}
{\cal{L}}({\bf p}) & = & {\cal{L}}_{g}(\log{M_{vir}}, c_{vir}, q; f_W) \nonumber \\
& \times & {\cal{L}}_{\mu}(\log{M_{vir}}, c_{vir}, q; f_W, \bar{n}_{\mu}, s_{eff}) \nonumber \\
& \times & {\cal{L}}_{{\cal{F}}}(\log{M_{vir}}, c_{vir}, q; f_W) \nonumber \\
& \times & {\cal{L}}_{{\cal{G}}}(\log{M_{vir}}, c_{vir}, q; f_W)
\label{eq: likeall}
\end{eqnarray}
where we have explicitely indicated which parameter each term depends on. Note that $(f_W, \bar{n}_{\mu}, s_{eff})$ are nuisance quantities which we one can marginalize over if only interested in constraining the cluster profile. Note that we will investigate under which conditions and to which extent flexion can help improving the mass (and concentration) determination. To this end, we will consider different data combination so that we will redefine the above likelihood by only retaining the corresponding terms.

A caveat is in order here. Eq.(\ref{eq: likeall}) implicitly assumes that the four different datasets (reduced shear, number counts, first and second flexion) are independent. While this is largely true for the shear and magnification, it is far less clear whether this actually comes true for shear and flexion too. Although there is still not a consensus on which method will be used to extract flexion from data, one can anticipate that shear and flexion measurements will rely on the same galaxies. As such, a certain correlation among errors should not be excluded. Indeed, \cite{VMB12} have shown that this is the case warning against the potential bias induced by neglecting such a correlation. However, two considerations come into help. First, the methods nowadays used for shear measurement such as, e.g., {\it lens}fit \citep{Lance07,Tom08,Lance12} or {\sc{im3shape}} \citep{Z13} are not suited for flexion so that one can anticipate that the two quantities will be estimated with different tools thus reducing the correlation among the errors. Second, as shown in \cite{R13}, given a galaxy sample suited for the shear measurement, only the subset with the largest S/N can be used to reliably estimate flexion since the noise on both $({\cal{F}}, {\cal{G}})$ severely increases as the galaxy luminosity decreases. As such, we do expect that, no matter which method is used, the samples used for the estimate of the shear and the flexion will be different (with faint galaxies dropping out from the second one) so that the correlation among the errors is largely reduced. Motivated by these two qualitative considerations and being not possible at the moment to guess a correlation matrix between $g$ and $({\cal{F}}, {\cal{G}})$ measurements, we will assume that shear and flexion are independent measurements so that the combined covariance matrix becomes block diagonal and we can rely on Eq.(\ref{eq: likeall}).

\section{Fisher matrix forecast}

In order to investigate whether and under which conditions the method proposed works in constraining the cosmological parameters, we perform a Fisher matrix analysis. The elements of the Fisher matrix  are given by

\begin{equation}
F_{ij} = \left . \frac{\partial^2 {\cal{L}}}{\partial p_i \partial p_j} \right |_{{\bf p} = {\bf p}_{fid}}\,,
\label{eq: deffish}
\end{equation}
being  $p_i$ the i\,-\,th parameter and ${\bf p}_{fid}$ the fiducial values. Since we have assumed that the different datasets are independent, it is easy to show that the total Fisher matrix simply reads (see, e.g., \citealt{Coe})

\begin{equation}
{\bf F} = {\bf F}_{g} + {\bf F}_{\mu}  + {\bf F}_{{\cal{F}}}  + {\bf F}_{{\cal{G}}}
\label{eq: sumfish}
\end{equation}
with ${\bf F}_X$ the Fisher matrix obtained by setting the likelihood ${\cal{L}}$ in Eq.(\ref{eq: deffish}) equal to that of the probe $X$. The inverse of the ${\bf F}$ gives the covariance matrix ${\bf C}$ whose $i$\,-\,th diagonal element is the minimum variance of the parameter $p_i$. When nuisance parameters are present, one should first marginalize over them in order to get a reduced matrix relative to the cluster quantities only, namely the log virial mass $\log{M_{vir}}$ and the concentration $c_{vir}$. This latter one is typically determined with such a large error that one simply marginalize over it too. However, we want here to check whether this is still the case when flexion is combined with shear and magnification so that we will also consider the error on $c_{vir}$.

\subsection{Fisher matrix and S/N ratio}

Inserting the likelihood definition for the proble $X$ (with $X = g, \mu, {\cal{F}}, {\cal{G}}$ depending on the dataset used) into Eq.(\ref{eq: deffish}) naively gives

\begin{eqnarray}
F_{ij} & = & \sum_{k = 1}^{{\cal{N}}_X}{\frac{1}{\sigma_{X}^{2}(\theta_k)}
\frac{\partial{X_{th}(\theta_k, {\bf p})}}{\partial{p_i}} \frac{\partial{X_{th}(\theta_k, {\bf p})}}{\partial{p_j}}} \nonumber \\
 & = & \sum_{k = 1}^{{\cal{N}}_X}{ \left [ \frac{X_{th}(\theta_k)}{\sigma_X(\theta_k)} \right ]^2
\frac{\partial{\ln{X_{th}(\theta_k)}}}{\partial{p_i}} \frac{\partial{\ln{X_{th}(\theta_k)}}}{\partial{p_j}}}
\label{eq: fmvssn}
\end{eqnarray}
which can be more conveniently expressed in terms of the observed quantities assuming that the estimator used in the measurement of $X_{th}$ is unbiased so that $X_{th}(\theta_k) = X_{obs}(\theta_k)$. In this case, the terms in square brackets in Eq.(\ref{eq: fmvssn}) is the signal\,-\,to\,-\,noise ratio for the measurement of $X$ in a bin centred on $\theta_k$. Splitting the error as the sum in quadrature of a statistical one $\varepsilon_X$ and an intrinsic scatter $\sigma_{X,int}$, we get

\begin{eqnarray}
\frac{X_{th}(\theta_k)}{\sigma_X(\theta_k)} & = & \frac{X_{obs}(\theta_k)}{\varepsilon_X(\theta_k)}
\left [ 1 + \frac{\sigma_{X,int}^2}{\varepsilon_X^2(\theta_k)} \frac{1}{N_X^2} \right ]^{-1/2} \nonumber \\
 & = & \nu_X(\theta_k; \log{M_{vir}}, c_{vir}, q; \sigma_{X,int}, n_g, \Delta_{\theta})
\label{eq: ston}
\end{eqnarray}
where $N_X$ is the number of galaxies in circular corona containing the systems used for the $X$ measurement, $n_g$ the source total number density,  and we have assumed that all bins have the same width $\Delta_{\theta}$. In the above relation, we have denoted with $\nu_X$ the overall S/N ratio of the measurement method after averaging over the sources in the given bin and used the fact that the intrinsic scatter contribution to the error scales with the inverse of the number of galaxies. Moreover, we have also explicitly listed the quantities the overall $S/N$ depends on to highlight its dependence on both the halo properties (which sets the amplitude of the signal to measure), the intrinsic scatter of the quantity, and the survey features. We can now rewrite the Fisher matrix elements for the probe $X$ as

\begin{equation}
F_{ij} = \sum_{k = 1}^{{\cal{N}}_X}{ \nu_{X}^{2}(\theta_k)\frac{\partial{\ln{X_{th}(\theta_k)}}}{\partial{p_i}} \frac{\partial{\ln{X_{th}(\theta_k)}}}{\partial{p_j}}}
\label{eq: fmvssnend}
\end{equation}
where one can again replaced $X_{obs}(\theta_k)$ with $X_{th}(\theta_k)$ In Eq.(\ref{eq: ston}) since, in the Fisher matrix approach, all the quantities are evaluated for the fiducial values of the parameters so that observed and theoretical quantities are equal.

In order to take care of the dependence of $\nu_X(\theta)$ on the halo parameters and its scaling with the position $\theta$, we proceed as follows. For fixed values of $(\log{M_{vir}}, c_{vir}, q, z)$ and given the observational quantities $(n_g, \Delta_{\theta})$, we first build a mock dataset as described later and, for each $\theta_k$ value, randomly extract $\sigma_X(\theta_k)/X_{th}(\theta_k)$ from a uniform distribution for $X = (g, {\cal{F}}, {\cal{G}})$. We check that the overall S/N for the shear is everywhere larger than the one for the flexion and then fit the profiles with a loglinear relation 

\begin{equation}
\log{\nu_X(\theta_k)} \simeq \nu_X^{(0)} + \nu_{X}^{(1)} \log{(\theta_k/2.0)}
\label{eq: nuxfit}
\end{equation}
which turns out to provide a reasonably good fit to the generated values. We repeat this procedure for many halo model parameters sets and, for each $q$, we approximate the above coefficients as 

\begin{equation}
\nu_{X}^{(n)} = a_{X}^{(n)} + b_{X}^{(n)} \log{ \left ( \frac{M_{vir}}{M_{piv}} \right )} 
+ c_{X}^{(n)} \log{\left ( \frac{1 + z}{1 + z_{piv}} \right )}
\label{eq: parscale}
\end{equation}
with $(\log{M_{piv}}, z_{piv}) = (14.5, 1.0)$ and the fit coefficients depending on $n = 1, 2$, the quantity of interest and the halo axial ratio $q$. Although the scatter of residuals can be non negligible, this procedure is nevertheless able to reproduce the correct scaling of the overall S/N $\nu_X$ with both the radial distance and the halo parameters so that we are confident that the Fisher matrix forecasts thus obtained are based on a realistic description of the measurement errrors.  

\subsection{Fiducial model}

Both the derivatives and the error term entering the Fisher matrix elements are computed setting the parameters to be constrained to fiducial values.

First, we have to set the halo properties, namely virial mass, concentration and redshift. The larger is the mass, the larger the signal will be so that one can naively expect that the $FoM$ and the constraints on $\log{M_{vir}}$ gets better and better as the mass increases. However, concentration also plays a role with larger $c_{vir}$ leading to a larger reduced shear. Moreover, for a given mass and concentration, the cluster redshift has to be taken into account too. Finally, we note that flattened haloes are expected to produce greater lensing signals than their spherical counterparts. In order to explore the parameter space, we will consider models with $\log{M_{vir}} = (14.0, 14.5, 15.0)$, $(z_d = 0.3, 0.9, 1.4)$, and $q = (0.75, 0.85, 0.95)$ and their different combinations. For each $(\log{M_{vir}}, z)$ combination, we set the concentration $c_{vir}$ using the $c_{vir}$\,-\,$M_{vir}$ relation of \cite{D08} neglecting its scatter. Note that weak lensing data are presently available for clusters mainly in the mass range $(1, 40) \times 10^{14} \ {\rm M_{\odot}}$ and redshift $0.2 \le z \le 1.4$ \citep{S15,GGS15} so that our choice is representative of actual samples. 

We have then to set the nuisance parameters $(f_W, \bar{n}_{\mu}, s_{eff})$. As already said, $f_W$ is expected to be of order unity so that we fix $f_W = 1.01$ stressing that its value has actually a negligible impact on the results. We then set $\bar{n}_{\mu} = 13.3 \ {\rm gal/arcmin}^2$ and $s_{eff} = 0.14$ which are typical values for CLASH clusters \citep{U13}. Note that these quantities can span a large range, but we do not explore the dependence of the results on them since we use magnification only as a reference being our main interest in flexion. For this same reason, we hold fixed $\nu_{\mu} = 3$ as for CLASH clusters although one can expect that space based observations could give a better S/N ratio.

The Fisher matrix results also depend on the intrinsic scatter for shear and flexion. We set $\sigma_{g,int} = 0.255$, while \cite{Leo07} recommended to use $\sigma_{{\cal{F}},int} a = 0.03$ and $\sigma_{{\cal{G}},int} a = 0.04$ with $a$ the typical angular size of the galaxy used to measure the flexion. Finally, we remind the reader that magnification is not affected by any intrinsic scatter so that it is $\sigma_{\mu,int} = 0$.

As shown in \cite{R13} from simulated data and in \cite{V11} from a first pioneering work on actual data, flexion is quite hard to measure so that the expected S/N is quite low. Moreover, only the highest S/N galaxies will allow a determination of $({\cal{F}}, {\cal{G}})$. We should therefore reduce the number density $n_g$ entering Eq.(\ref{eq: fmvssnend}) by a factor $\epsilon_f$ to account for this effect. However, given that we have absorbed $n_g$ into the definition of the overall S/N ratio and already imposed that $\nu_g > \nu_{{\cal{F}}}, \nu_{{\cal{G}}}$, we do not include $\varepsilon_f$ as a parameter taking its value fixed to unity.

\section{Results}

The last ingredient we need to compute the Fisher matrix is the distribution of the data, i.e. what is the radial range probed and in how many bins it is divided. Following again \cite{U13}, we consider the range $0.2 \le R/{\rm arcmin} \le 16$ thus excluding the very inner regions which are in the strong lensing regime and the outer ones where the second halo term dominates. We divide this range in 10 logarithmically spaced bins and set $n_g = 30 \ {\rm gal/arcmin^2}$ as expected for the Euclid satellite \citep{RB}.

In all the results presented below, we will assume that the nuisance parameters $(f_W, \bar{n}_{\mu}, s_{eff})$ are fixed to their fiducial values since they can be easily be estimated from independent observations. We are therefore left with the three parameters $(\log{M_{vir}}, c_{vir}, q)$. The constraints on each one of them is obtained by marginalizing over the remaining two.

\begin{figure*}
\resizebox{\hsize}{!}
{\includegraphics[width=4.5cm]{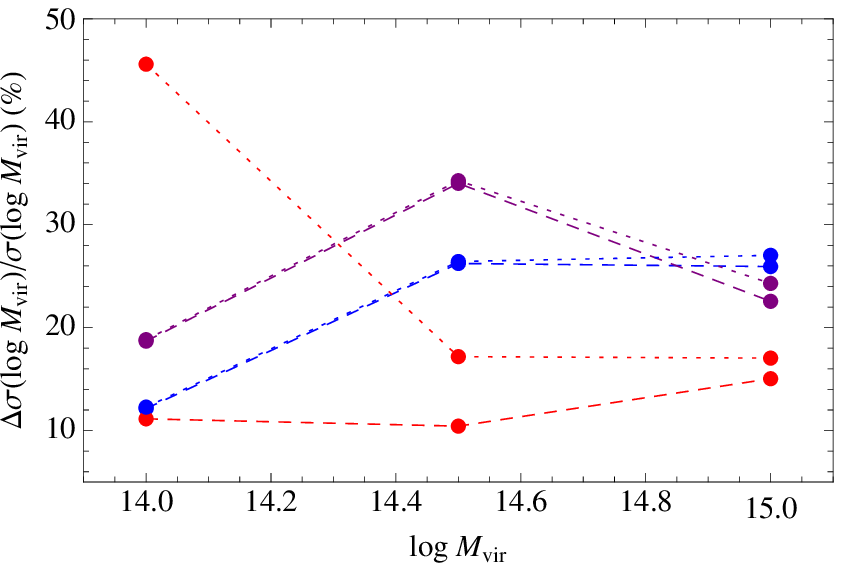}
\includegraphics[width=4.5cm]{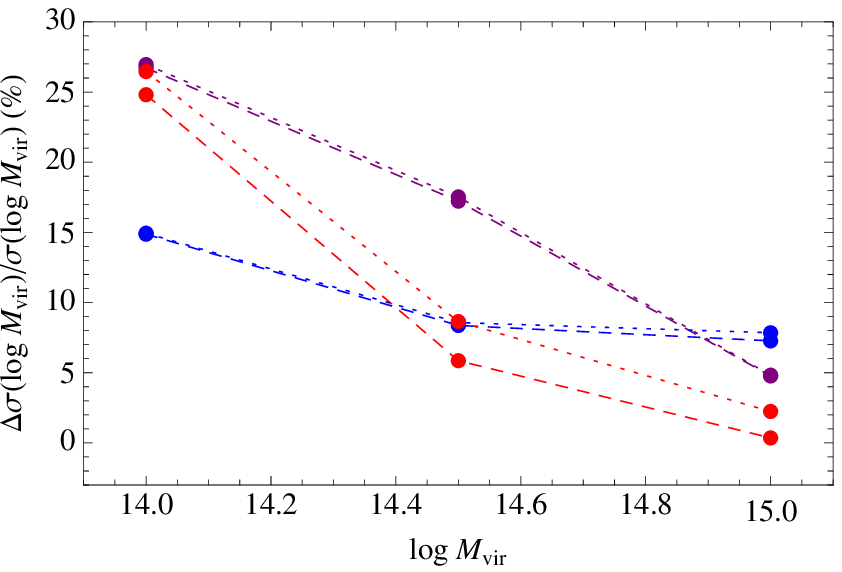}
\includegraphics[width=4.5cm]{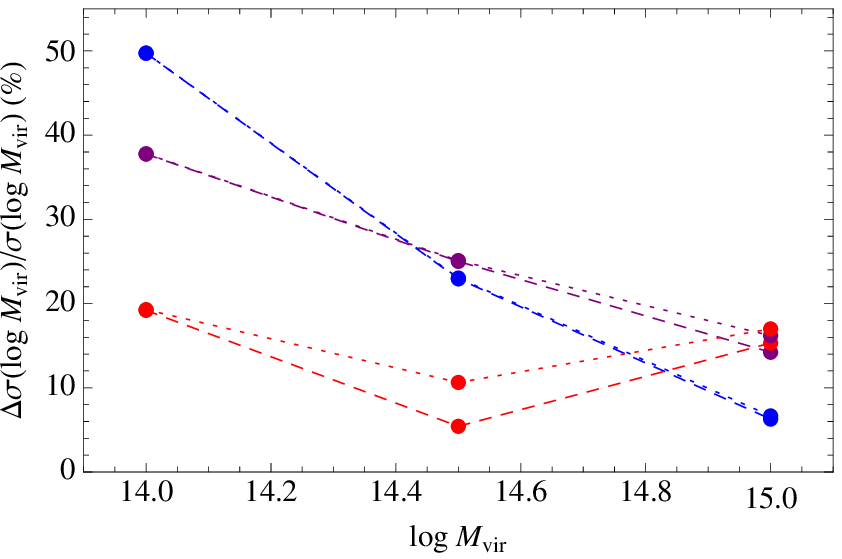}}
\resizebox{\hsize}{!}
{\includegraphics[width=4.5cm]{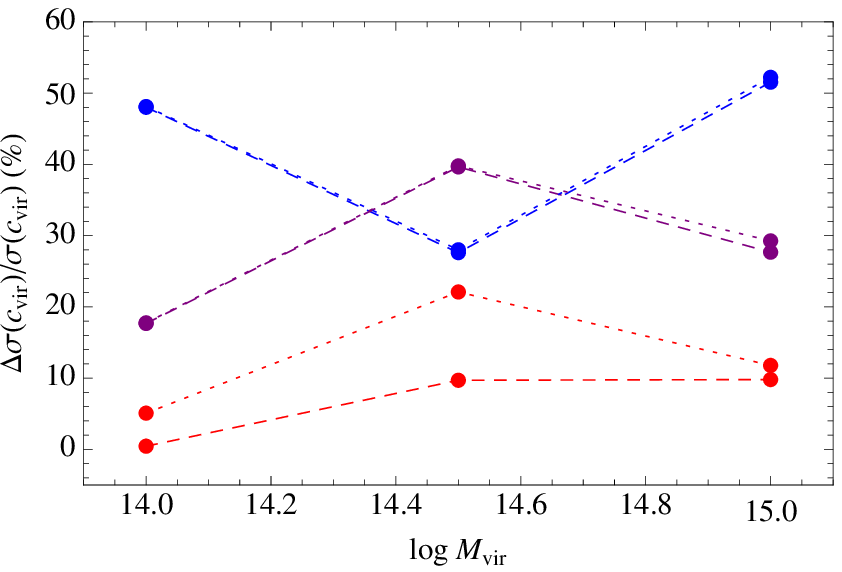}
\includegraphics[width=4.5cm]{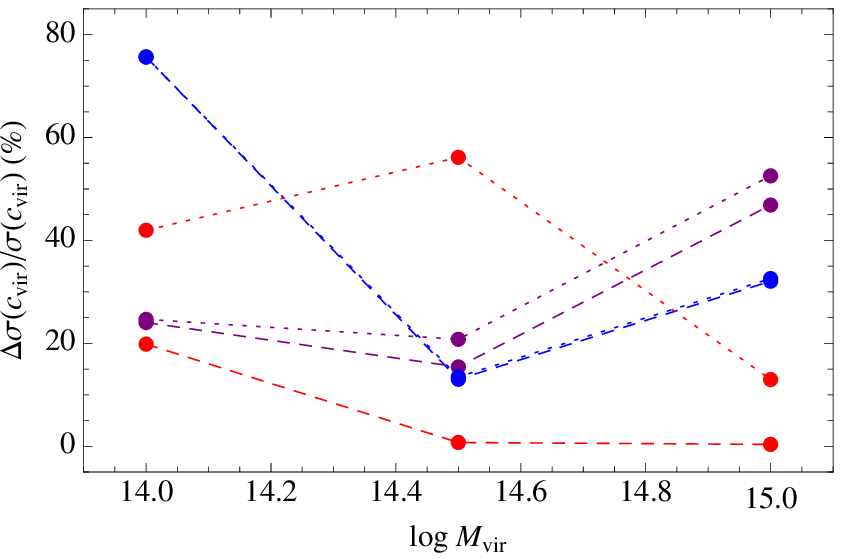}
\includegraphics[width=4.5cm]{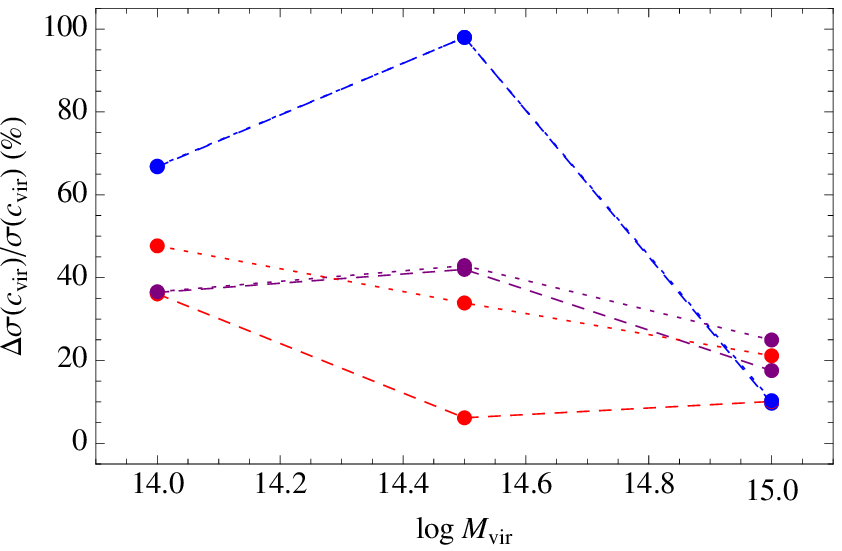}}
\caption{Relative improvement in error on mass $\mu$ (top panels) and concentration $c_{vir}$ (bottom panels) with respect to the shear only constraints for clusters at redshift $z = 0.3$ (left), $z = 0.9$ (centre), $z = 1.4$ (right). Dashed (dotted) lines refer to the case with shear\,+\,flexion (shear\,+\,flexion\,+\,magnification), while blue, purple, red colors denote models with $q = (0.75, 0.85, 0.95)$, respectively.}
\label{fig: smucvrationow}
\end{figure*}

\subsection{Present day constraints}

We first consider the case with errors on shear and flexion mimicking the present day ones investigating to which extent flexion can improve the constraints on mass and concentration in conjuction with magnification or replacing magnification itself. Let us first discuss the results on the mass $\mu = \log{M_{vir}}$. The relative error $\sigma_{\mu}/\mu$ from shear\,+\,magnification only turns out to be

\begin{displaymath}
\sigma_{\mu}(z = 0.3) = \left \{
\begin{array}{ll}
(0.031, 0.031, 0.018) & {\rm for} \ \mu = 14.0 \\
(0.025, 0.022, 0.019) & {\rm for} \  \mu = 14.5 \\
(0.023, 0.016, 0.016) & {\rm for} \  \mu = 15.0 \\
\end{array}
\right . \ ,
\end{displaymath}

\begin{displaymath}
\sigma_{\mu}(z = 0.9) = \left \{
\begin{array}{ll}
(0.046, 0.045, 0.043) & {\rm for} \ \mu = 14.0 \\
(0.030, 0.030, 0.018) & {\rm for} \  \mu = 14.5 \\
(0.026, 0.005, 0.010) & {\rm for} \  \mu = 15.0 \\
\end{array}
\right . \ ,
\end{displaymath}

\begin{displaymath}
\sigma_{\mu}(z = 1.4) = \left \{
\begin{array}{ll}
(0.084, 0.064, 0.041) & {\rm for} \ \mu = 14.0 \\
(0.039, 0.038, 0.018) & {\rm for} \  \mu = 14.5 \\
(0.028, 0.029, 0.030) & {\rm for} \  \mu = 15.0 \\
\end{array}
\right . \ ,
\end{displaymath}
where the different values for each row refers to a cluster with that mass and redshift and $q = (0.75, 0.85, 0.95)$, respectively. Note that these values are indeed quite optimistic if compared to the results from actual cluster mass determination (see, e.g., \citealt{Clash14}). This is partly due to the larger number density and smaller intrinsic shear variance and partly to the well known fact that Fisher matrix variances are lower limits to the errors one can achieve. Actually, such small errors are mainly due to having assumed a diagonal covariance matrix and not included either systematics errors and the contribution from the large scale structure which is dominant for small signals and in the outer cluster regions. However, we are here mainly interested in comparing how errors change when including flexion rather than forecasting precise values. Set in other words, we concern here with ratios of errors, not absolute values. Since the neglected terms do not depend on including or not flexion, we are confident that the results presented below qualitatively hold even if the full error analysis is carried on. 

Top panels in Fig.\,\ref{fig: smucvrationow} shows how the constraints on the halo mass improves when flexion instead of magnification is combined with shear. In particular, we plot $\Delta \sigma(\mu, {\cal{O}})/\sigma(\mu, WL)$ with $\sigma(X, {\cal{O}})$ the error on the quantity $X$ when using the observable\footnote{Hereafter, we use ${\cal{O}} = WL, WL\,+\,Flex, All$ to refer to shear only, shear\,+\,flexion, shear\,+\,flexion\,+\,magnification.} ${\cal{O}}$ and $\Delta \sigma = \sigma(X, WL) - \sigma(X, {\cal{O}})$. A positive value means that the use of the observable set ${\cal{O}}$ has reduced the error on $X$ by a factor $\Delta \sigma/\sigma$. 

\begin{figure*}
\resizebox{\hsize}{!}
{\includegraphics[width=4.5cm]{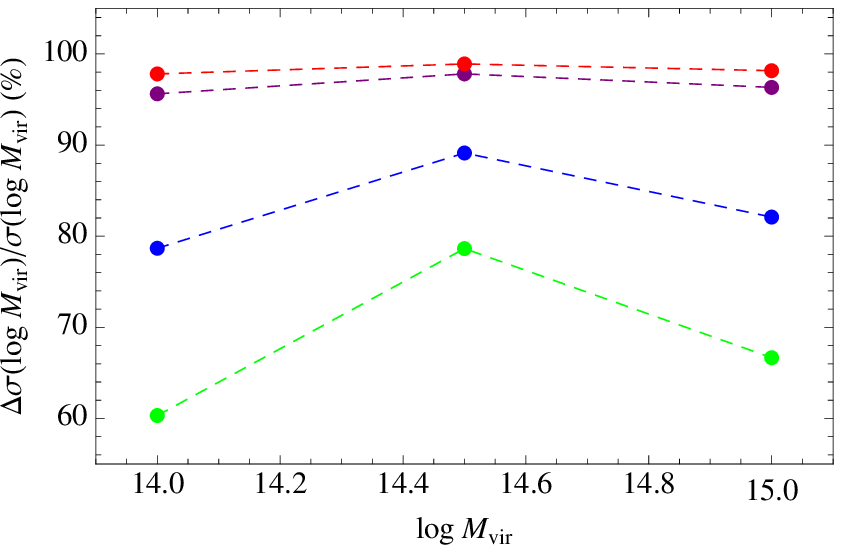}
\includegraphics[width=4.5cm]{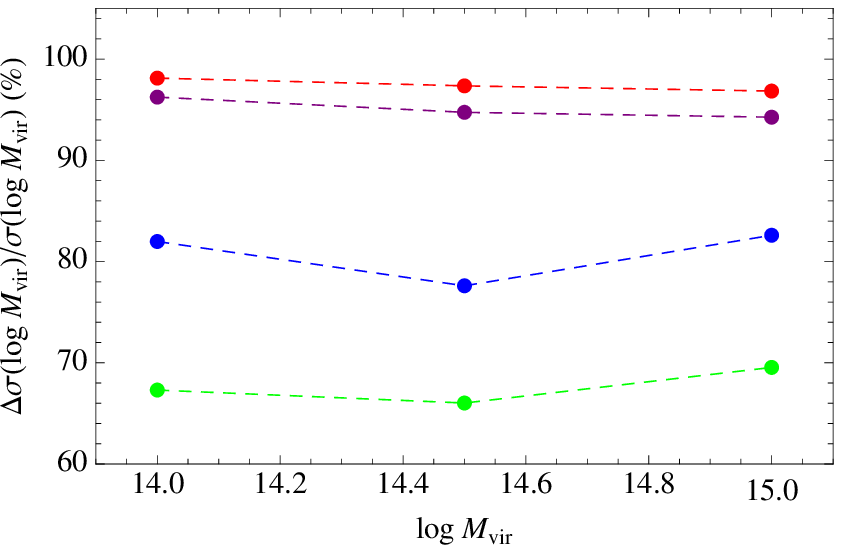}
\includegraphics[width=4.5cm]{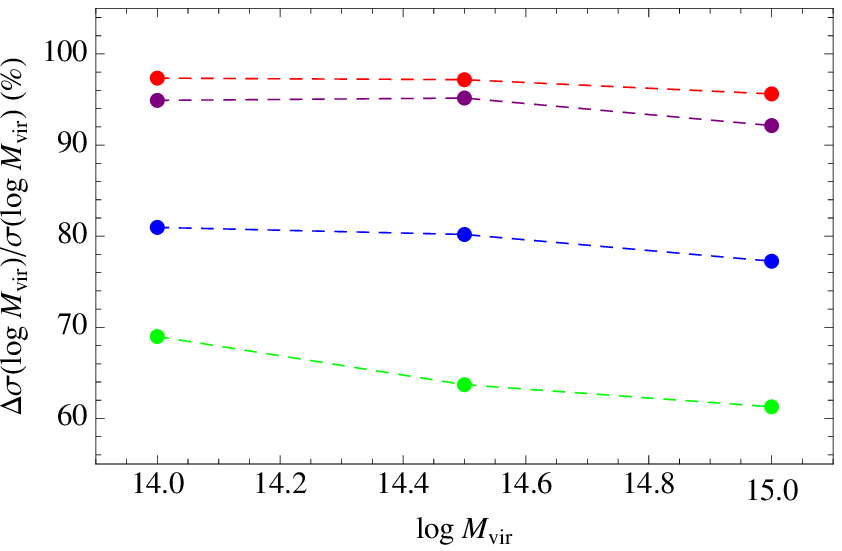}}
\resizebox{\hsize}{!}
{\includegraphics[width=4.5cm]{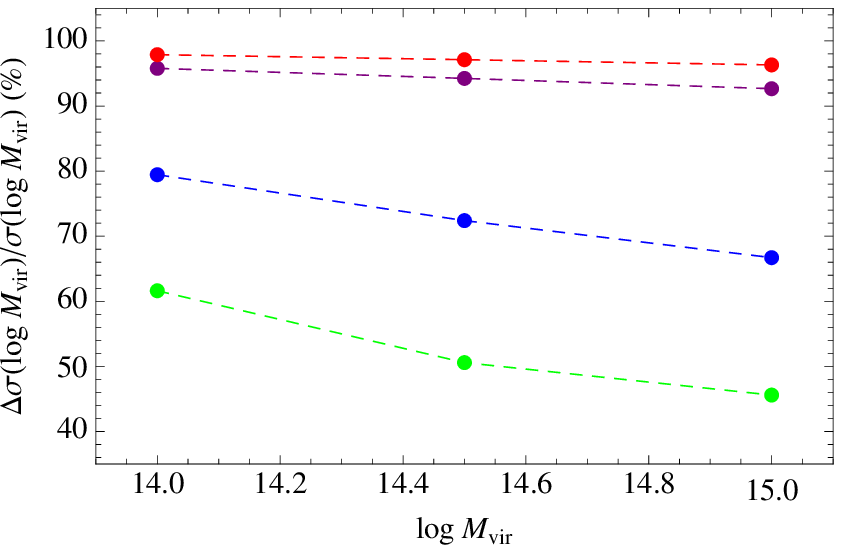}
\includegraphics[width=4.5cm]{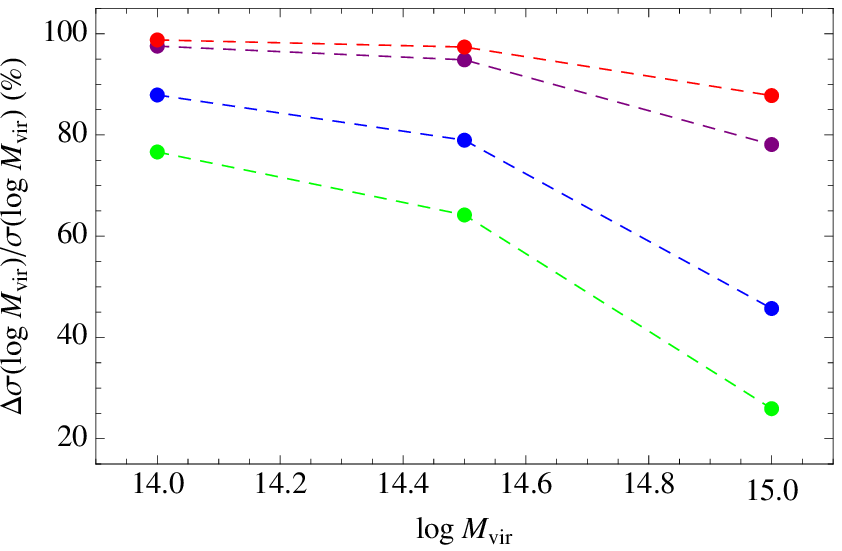}
\includegraphics[width=4.5cm]{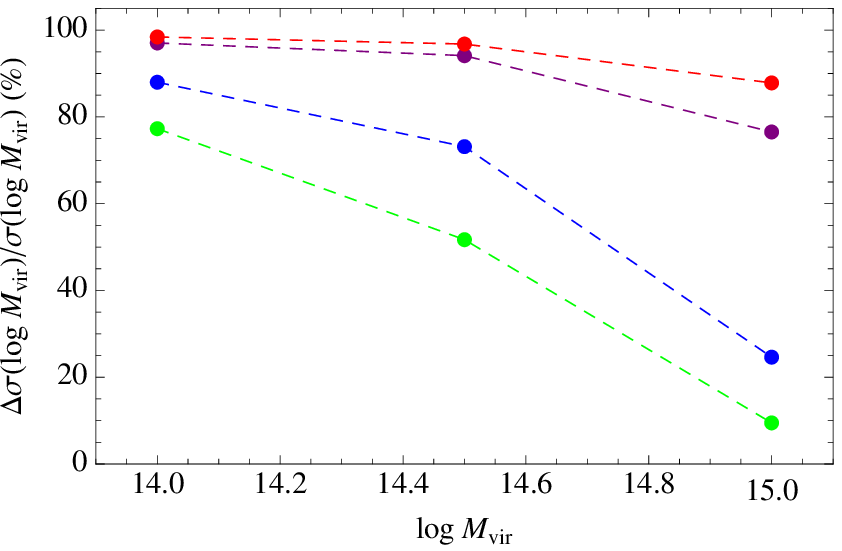}}
\resizebox{\hsize}{!}
{\includegraphics[width=4.5cm]{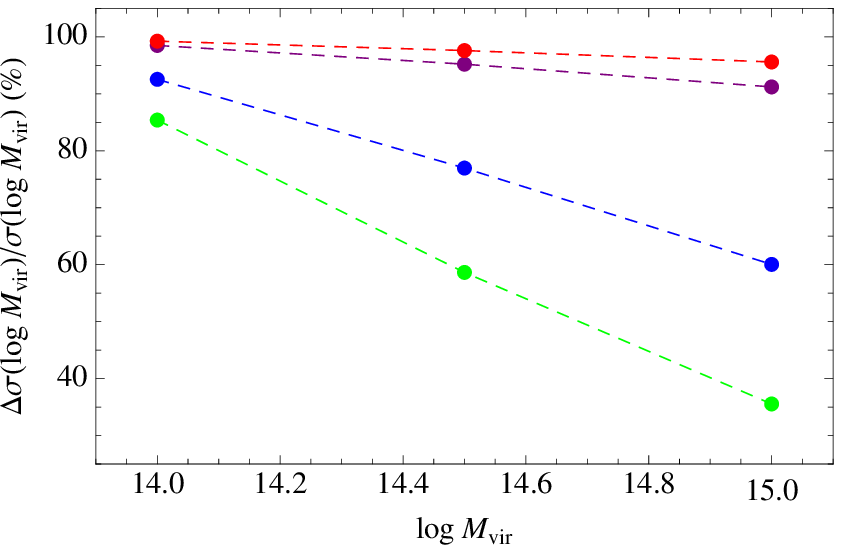}
\includegraphics[width=4.5cm]{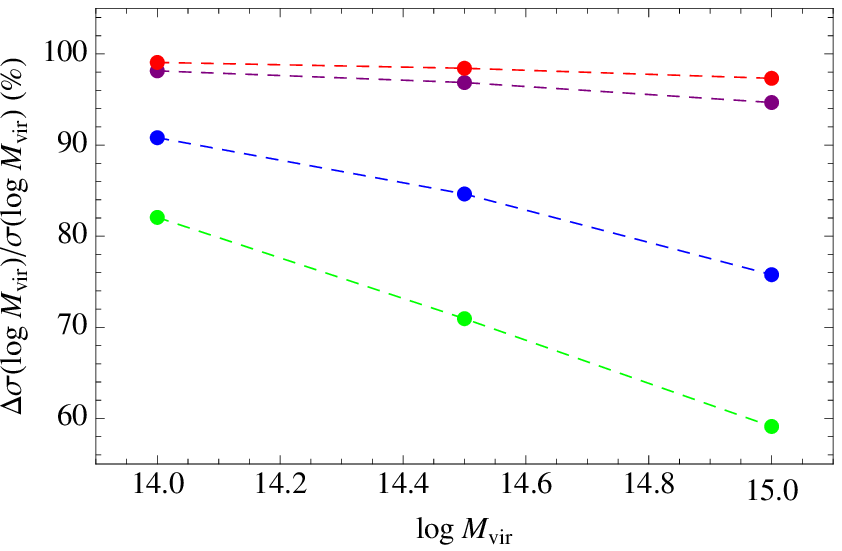}
\includegraphics[width=4.5cm]{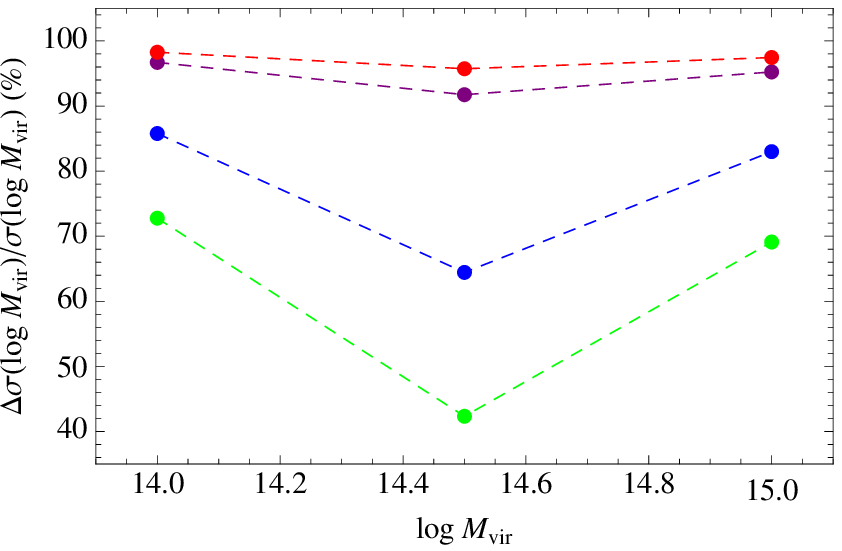}}
\caption{Relative improvement in error on mass $\mu$ with respect to the shear only constraints for clusters at redshift $z = 0.3$ (top), $z = 0.9$ (centre), $z = 1.4$ (bottom). Green, blue, purple, red lines refer to the case with the flexion S/N increased by a factor $b_{{\cal{F}}} = (5, 10, 50, 100)$, while $q = (0.75, 0.85, 0.95)$ from left to right.}
\label{fig: smuratiobgsetbfchange}
\end{figure*}

Some qualitative lessons can be drawn from Fig.\,\ref{fig: smucvrationow}. The first striking conclusion is that flexion can indeed boost the accuracy in mass determination by a significative factor even when the cluster is at high redshift. Second, we note that adding magnification to shear and flexion does not improve the constraints as can be seen noting that dashed and dotted lines are almost superimposed unless nearly spherical haloes($q = 0.95$) are considered. In this case, the flexion signal is smaller so that the impact of magnification can be better appreciated. As a roughly general result, the boost in the mass determination accuracy is larger for smaller $q$ in agreement with the qualitative expectation that, for given cluster mass and redshift, both shear and flexion are larger for more flattened haloes. However, some inversion is sometimes possible depending on the $(\mu, z)$ values. Indeed, inferring general trends is far from trivial. On the one hand, one could expect that $\Delta \sigma/\sigma$ is larger for more massive haloes and lower redshift clusters. Actually, the Fisher matrix elements depend on both the derivative of the shear and flexion with respect to the model parameters and the S/N ratios of the quantities to be fitted which again depend on the model parameters. As a consequence, non monotonic trends with $(\mu, z, q)$ are possible as can be read from the plots.

Let us now consider the constraints on the concentration $c_{vir}$. Using shear and magnification only leads to

\begin{displaymath}
\sigma_{c}(z = 0.3) = \left \{
\begin{array}{ll}
(0.38, 0.12, 0.03) & {\rm for} \ \mu = 14.0 \\
(0.24, 0.10, 0.02) & {\rm for} \  \mu = 14.5 \\
(0.18, 0.05, 0.04) & {\rm for} \  \mu = 15.0 \\
\end{array}
\right . \ ,
\end{displaymath}

\begin{displaymath}
\sigma_{c}(z = 0.9) = \left \{
\begin{array}{ll}
(0.33, 0.24, 0.08) & {\rm for} \ \mu = 14.0 \\
(0.24, 0.14, 0.02) & {\rm for} \  \mu = 14.5 \\
(0.24, 0.04, 0.10) & {\rm for} \  \mu = 15.0 \\
\end{array}
\right . \ ,
\end{displaymath}

\begin{displaymath}
\sigma_{c}(z = 1.4) = \left \{
\begin{array}{ll}
(1.02, 0.43, 0.29) & {\rm for} \ \mu = 14.0 \\
(0.46, 0.34, 0.07) & {\rm for} \  \mu = 14.5 \\
(0.18, 0.09, 0.02) & {\rm for} \  \mu = 15.0 \\
\end{array}
\right . \ ,
\end{displaymath}
where the different values for each row refers to a cluster with that mass and redshift and $q = (0.75, 0.85, 0.95)$, respectively. Although the constraints are again optimistic, we are nevertheless more interested in the results shown in the lower panels of Fig.\,\ref{fig: smucvrationow} where the improvement in accuracy due to adding flexion to shear is plotted for different combinations of mass, redshift and axial ratio. Roughly, we find the same qualitative results as for the mass, but the amplitude of $\Delta \sigma(c_{vir}, {\cal{O}})/\sigma(c_{vir}, WL)$ is definitively larger. This is a consequence of flexion breaking the mass\,-\,concentration degeneracy so that the constraints on $c_{vir}$ greatly improve. This makes magnification not quite helpful unless the halo is nearly spherical in which case the flexion signal is greatly reduced. Somewhat surprisingly, $\Delta \sigma/\sigma$ is larger for larger $z$ as can be appreciated looking at the ordinate scale. This is actually a consequence of how $\Delta \sigma/\sigma$ is defined. Indeed, at large $z$, the errors on $c_{vir}$ for given $\mu$ are larger for larger $z$, but so is $\Delta \sigma$ thus explaining why flexion is more effective for higher redshift clusters.

\subsection{Future constraints}

The above results convincingly show that it is worth combining flexion and shear to improve the accuracy in mass and concentration determination. We therefore wonder here whether it is possible to do still better by increasing the S/N in the measurement of shear and/or flexion. To this end, we repeat the above analysis replaing $\nu_{X}$ in Eq.(\ref{eq: fmvssn}) with $b_X \nu_X$, i.e., we assume that the S/N for the probe $X$ is increased by a factor $b_X$ thanks to a future advance in measurment techniques (because of better quality data and/or progress in detection algorithms). 

As a first case, we set $b_g = 1$, $b_{{\cal{F}}} = b_{{\cal{G}}}$, and show in Fig.\,\ref{fig: smuratiobgsetbfchange} the boost in mass determination accuracy for different $(\mu, z, q)$ combinations and $b_{{\cal{F}}}$ values. A naive look at the plots is enough to convince that it is worth investing time to improve the flexion S/N. For instance, for a cluster with $(\mu, z, q) = (14.0, 0.3, 0.75)$, a factor $5$ increase in $\nu_{{\cal{F}}}$ makes $\Delta \sigma(\mu, All)/\sigma(\mu, WL)$ increase from 12 to $60\%$. However, note that the scaling of $\Delta \sigma/\sigma$ is not linear with $b_{{\cal{F}}}$, but rather quickly saturates so that it is actually not worth trying to get $b_{\cal{F}} > 50$. Qualitatively similar results also hold for concentration so that we do not report here the corrisponding plots. 

\begin{figure*}
\resizebox{\hsize}{!}
{\includegraphics[width=4.5cm]{smuratiozd03bg01q85.eps}
\includegraphics[width=4.5cm]{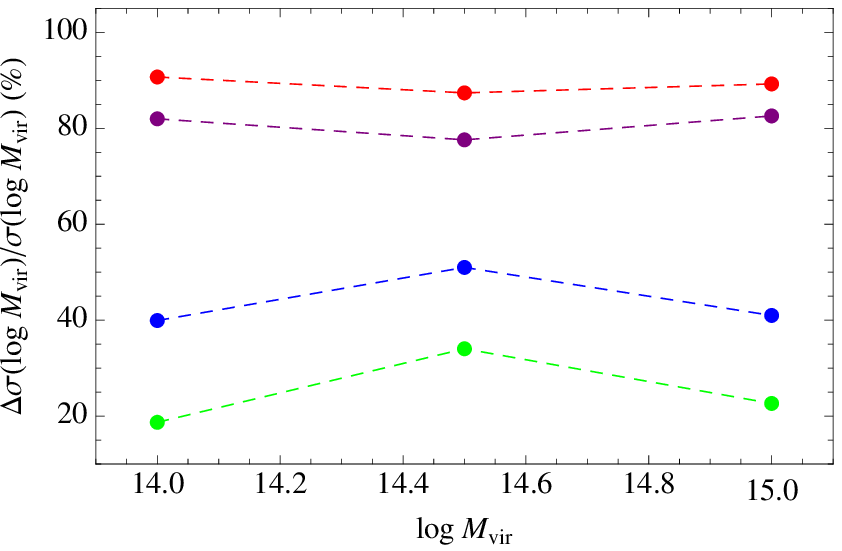}
\includegraphics[width=4.5cm]{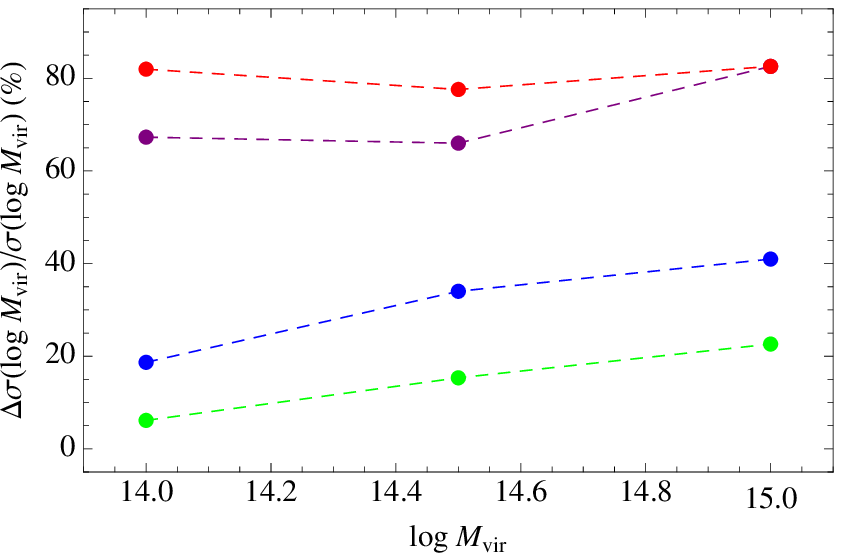}}
\resizebox{\hsize}{!}
{\includegraphics[width=4.5cm]{smuratiozd09bg01q85.eps}
\includegraphics[width=4.5cm]{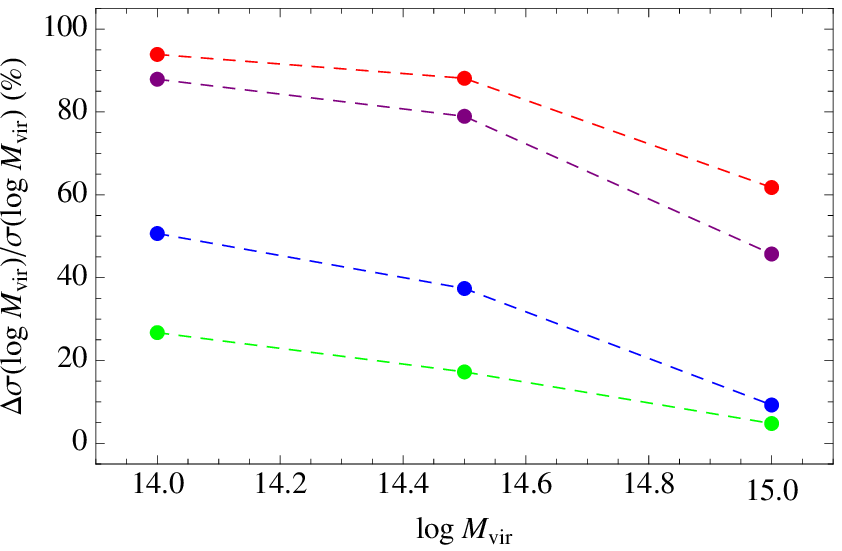}
\includegraphics[width=4.5cm]{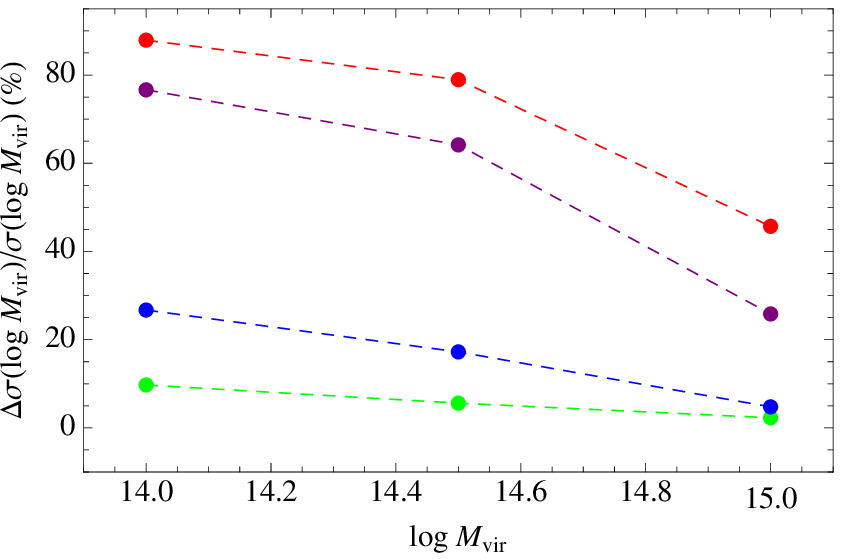}}
\resizebox{\hsize}{!}
{\includegraphics[width=4.5cm]{smuratiozd14bg01q85.eps}
\includegraphics[width=4.5cm]{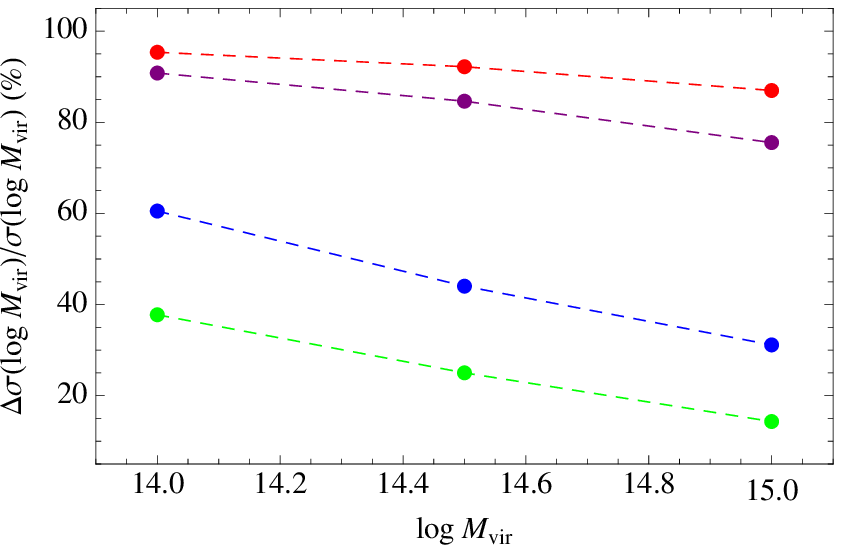}
\includegraphics[width=4.5cm]{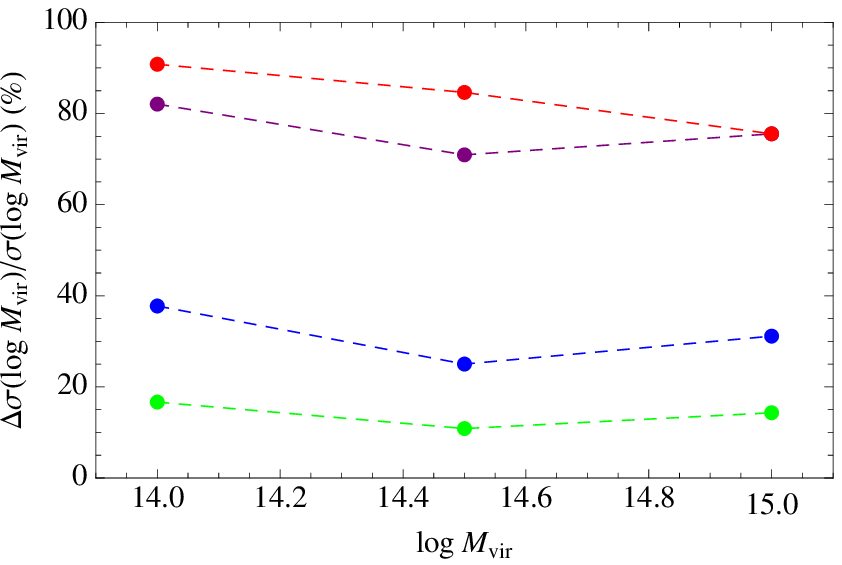}}
\caption{Relative improvement in error on mass $\mu$ with respect to the shear only constraints for clusters at redshift $z = 0.3$ (top), $z = 0.9$ (centre), $z = 1.4$ (bottom) and with axial ratio $q = 0.85$. Green, blue, purple, red lines refer to the case with the flexion S/N increased by a factor $b_{{\cal{F}}} = (5, 10, 50, 100)$, while $b_g = (1, 5 ,10)$ from left to right.}
\label{fig: smuratiobgchangebfchange}
\end{figure*}

Although shear measurement is definitely a more mature field than flexion measurement, it is nevertheless unlikely that future devolopements will improve $\nu_{{\cal{F}}}$ only, while leaving $\nu_{g}$ unchanged. Fig.\,\ref{fig: smuratiobgchangebfchange}  is therefore the same as Fig.\,\ref{fig: smuratiobgsetbfchange}, but now we fix the halo axial ratio and explores the dependence on $b_g$. It turns out that the larger is $b_g$, the more one should improve $\nu_{{\cal{F}}}$ in order to have a significant reduction of the error on the mass due to the use of flexion. As a consequence, the same improvement in mass determination accuracy can be obtained by keeping unchanged the flexion S/N, but increasing the shear one. 

\section{Conclusions}

The search for an efficient way to weight galaxy clusters dates back to the first understanding of their relevant role in constraining cosmological parameters and structure formation scenarios. As far as first shear measurements became possible, weak lensing has emerged as valid tool to achieve this goal thanks to its ability to probe the matter content no matter whether the cluster is relaxed or not and its state of thermal equilibrium. Although cluster masses through weak lensing have been measured for hundreds of clusters (see, e.g., the compilation in \citealt{S15} and refs. therein), there is yet room for improvement being the typical uncertainty on the mass still far from negligible. Flexion has been proposed as a possible way to boost mass measurement accuracy so that we have here investigated under which conditions and to which extent this is indeed the case. 

Our Fisher matrix analysis convicingly show that the use of flexion can indeed reduce the error on the mass determination by a significative amount under realistic assumption on the overall S/N ratio of $(g, {\cal{F}}, {\cal{G}})$. The gain in accuracy is larger for more flattened haloes as a result of the larger flexion signal, while the scaling with the cluster mass and redshift is non monotonic as a consequence of the complicated dependence of the Fisher matrix elements on both the derivative of the observed quantities and the way the S/N ratio scale with the halo parameters. We have also explored how these results change if future developments in shear and/or flexion measurement codes allow to improve the overall S/N. It turns out that the gain in accuracy saturates with the flexion S/N so that investigating efforts to improve $\nu_{{\cal{F}}}$ more than a factor $\sim 50$ is a non rewarding strategy. Moreover, Fig.\,\ref{fig: smuratiobgchangebfchange} shows that pushing up the flexion S/N has a lower and lower impact as shear S/N gets larger and larger. However, while there is a wide room for ameliorating flexion measurements techniques (being this field still at its infancy), it is unlikely that order of magnitude improvements in the shear S/N are possible in the near future so that the use of flexion will likely be a valid help to narrow down the mass confidence range.

Although quite encouraging, our results need to be furtherly tested since they are based on some reasonable yet qualitative assumptions. First, we have combined shear, magnification and flexion by simply defining the full likelihood function (whose derivatives enter the Fisher matrix) as the product of the three single likelihood functions. This is the same as assuming that the errors on the three probes are uncorrelated. While this is true for shear and magnification, the question is still open for shear and flexion. Indeed, \cite{VMB12} pointed out a correlation between shear and flexion noise which can bias flexion measurements if not correctly taken into account. However, the same authors proposed a method to reduce both the bias and the correlation, but a precise accounting of this effect in a Fisher matrix analysis asks for a modeling of the covariance matrix which is still unavailable at the moment. Working out such a model asks for an empirical knowledge of the flexion noise from image simulations and mock measurements so that we have here assumed the ideal situation that future methods will be able to reduce the correlation among these two different probes of the lensing potential. Should this not be the case, the Fisher matrix analysis must be repeated including the relevant cross talk terms possibly reducing the boost in mass accuracy we find out here. It is, however, worth noting that the increase in accuracy we have forecast is so large that we are confident flexion will still stand out as useful complement to shear for cluster mass determination.  

A second simplifyng assumption concerns the dependence of the shear and flexion S/N on the radial distance from the cluster centre. The lack of any hint on the functional expression of $\nu_X(R)$ (with $X = g, {\cal{F}}, {\cal{G}}$ for shear, first and second flexion, respectively) has forced us to work out a reasonable procedure to find out a scaling with both the $R$ and the halo parameters. However, given the key role of these quantities in the forecast analysis, it is worth readdressing this issue in more detail. Real and mock data can give us a hint on how to model $\nu_g(R)$, while much more work is needed for $\nu_{{\cal{F}}}(R)$ and $\nu_{{\cal{G}}}(R)$ being flexion measurement method still to be decided. 

As a concluding comment, we remark that our adimittedly preliminary and based on simplified treatment of errors results nevertheless strenghten the idea that flexion can represent the ideal complement to shear in solving the weighting the clusters problem. Although working out algorithms for flexion measurements with sufficiently high S/N can be a tremendously difficult task, the promise of reducing the mass error makes it definitely worth the trouble. 

 \section*{Acknowledgments}

VFC thanks M. Sereno and S. Pilo for interesting discussions. The authors acknowledge D. Bacon, H. Hoekstra and T. Kitching for comments on a preliminary version of the manuscript, and an anonymous referee for his/her suggestions that have led us to revise the paper. VFC and XE are funded by Italian Space Agency (ASI) through contract Euclid\,-\,IC (I/031/10/0) and acknowledge financial contribution from the agreement ASI/INAF/I/023/12/0.

\appendix

\section{Impact of substructures}

The derivatives of the reduced shear, magnification and flexion enter the elements of the Fisher matrix thus asking for a halo model in order to compute these quantities. We have adopted an elliptical NFW profile thus implicitly assuming that the cluster is a single component system. Actually, this is not the case because of the presence of substructures so that the total convergence should be written as

\begin{equation}
\kappa(R) = \kappa_{cl}(R) + \sum_{i = 1}^{{\cal{N}}_s}{\kappa_i(R - R_{ci})}
\label{eq: kappatot}
\end{equation}
where $\kappa_{cl}$ ($\kappa_i$) is the cluster ($i$\,-\,th substructure) contribution to the total convergence and we have set $q = 1$ for both the cluster and its ${\cal{N}}_s$ substructures. In the weak field limit, the lensing is additive so that we can compute the total shear and flexion profiles by simply adding the contribution of the cluster and its substrucures. We model all contributors as spherical NFW system and, for given values of the cluster parameters $(\log{M_{vir}}, z)$, set the concentration according to the \cite{D08} relation. Following \cite{Moka} and refs. therein, we proceed as follows to set the substructure parameters.

\begin{enumerate}
\item{We set the distance of the substructure distance $R_{ci}$ from the halo centre (taken as origin of the coordinate system) sampling from the cumulative distribution function

\begin{displaymath}
\frac{n(< x = R/R_s)}{{\cal{N}}_s} = \frac{(1 + \alpha_s c_{vir}) x^{\beta_s}}{1 + \alpha_s c_{vir} x^2}
\end{displaymath}
with $(\alpha_s, \beta_s) = (0.244, 2.75)$.} \\

\item{We assign to the substructure a mass $m_{vir}$ sampling from the subhalo mass function approximated\footnote{We have here removed the term $\bar{c}/c_{vir}$ which originates from the scatter in the $M_{vir}$\,-\,$c_{vir}$ relation that we have neglected.} as

\begin{displaymath}
\frac{1}{M_{vir}} \frac{dN_m}{dm_{vir}} = A (1 + z)^{1/2} m^{\alpha_m} 
\exp{\left [ - \beta_m \left ( \frac{m_{vir}}{M_{vir}} \right )^3 \right ]}
\end{displaymath}
where $(M_{vir}, z)$ refer to the parent halo and it is $(A, \alpha_m, \beta_m) = (9.33 \times 10^{-4}, -0.9, 12.2715)$.} \\

\item{We use the \cite{D08} mass\,-\,concentration relation to set the substructure $c_{vir}$ value.} \\

\item{We repeat steps (i)\,-\,(iii) until the total substructure mass is a fraction $f_{sub}$ of the total mass.} \\

\end{enumerate}
Once the halo and substructure parameters have been set, we compute the shear and flexion profiles for each of the bin used in the fitting procedure. As expected, the susbstructure contribution is quite negligible for the inner bins where the typically one to two orders of magnitude larger mass of the cluster makes the corresponding terms in the sum greatly dominant. The situation can be reversed in the outer bins where the cluster contribution is vanishing. Should a substructure be present close to the cluster virial radius, its effect can become dominant. However, in these bins, the shear and flexion signals are so small that are hardly measurable at all. Indeed, we have checked that, for values of $f_{sub} \sim few\%$, the systematic error made by neglecting the substructure contribution is smaller than the statistical one. We therefore tentatively conclude that substrcutures can be neglected, but we caution the reader that such a conclusion could be revised should flexion measurement algorithms significantly improve statistical accuracy.

\end{document}